\documentclass[aps,prc,twocolumn,superscriptaddress]{revtex4}

\usepackage{epsfig}
\begin{document}

\title{\Large Examination of level density prescriptions in the interpretation of high energy $\gamma$-ray spectra}
\author{Srijit Bhattacharya}
\affiliation{Department of Physics, Barasat Govt. College, Barasat, N 24 Pgs, Kolkata-700124, India }

\author{Deepak Pandit}
\affiliation{Variable Energy Cyclotron Centre, 1/AF, Bidhannagar, Kolkata-700064, India}

\author{Balaram Dey}
\affiliation{Variable Energy Cyclotron Centre, 1/AF, Bidhannagar, Kolkata-700064, India}

\author{Debasish Mondal}
\affiliation{Variable Energy Cyclotron Centre, 1/AF, Bidhannagar, Kolkata-700064, India}

\author{S. Mukhopadhyay}
\affiliation{Variable Energy Cyclotron Centre, 1/AF, Bidhannagar, Kolkata-700064, India}

\author{Surajit Pal}
\affiliation{Variable Energy Cyclotron Centre, 1/AF, Bidhannagar, Kolkata-700064, India}

\author{A. De}
\affiliation{Department of Physics, Raniganj Girls' College, Raniganj-713358, India}

\author{S. R. Banerjee}
\email[e-mail:]{srb@vecc.gov.in}
\affiliation{Variable Energy Cyclotron Centre, 1/AF, Bidhannagar, Kolkata-700064, India}


\date{\today}

\begin{abstract}
High energy $\gamma$-ray spectra measured by our group involving the compound nuclei (CN) $^{63}$Cu at 
excitation energy ($E^*$) $\sim$ 36 MeV with average angular momentum ($J$) = 12 - 17 $\hbar$, $^{97}$Tc
at $E^* \sim$ 29 - 50 MeV with $J$ = 12 - 14 $\hbar$, $^{113}$Sb at $E^*$ = 109 MeV and 121 MeV with 
$J$ = 49 - 59 $\hbar$ and $^{201}$Tl at $E^*$ = 39.5, 47.5 MeV with $J$ = 18 - 24 $\hbar$ have been analyzed 
utilizing the level density prescriptions of (i)Ignatyuk, Smirenkin and Tishin (IST), (ii)Budtz-Jorgensen 
and Knitter (BJK), and (iii) Kataria, Ramamurthy and Kapoor (KRK). 
These three prescriptions have been tested for correct statistical model 
description of high energy $\gamma$-rays in the light of extracting the giant dipole resonance (GDR) 
parameters at low excitation energy and spin where shell effects might play an important role as well as at 
high excitation energy where shell effects have melted. Interestingly, only the IST level density prescription
could explain the high energy $\gamma$-ray spectra with reasonable GDR parameters for all the four nuclei.

\end{abstract}

\maketitle
\section{Introduction}
One of the most exciting topics in contemporary nuclear physics is the study of nuclear structure under extreme conditions of nuclear temperature ($T$) and angular momentum ($J$). The giant dipole resonance (GDR), an archetypical example of collective vibrational mode built on excited nuclear states, provides us the insight of exotic nuclear shapes and structure \cite{gaa1, sno1, haa1}. Ardent experimental and theoretical interests \cite{alh1, Kic87, bra1, Bor91, orm1, Kus98, kel1, Maj04, Sri08, Drc10, dipu1, supm11, dipu2, dipu3, dipu4, bala,heck03, Dang11, dipu6} can be seen over the years in studying the properties of the GDR built on excited states in nuclei as this collective mode is strongly coupled to nuclear damping and shape degrees of freedom. The strength ($S_{GDR}$), centroid energy ($E_{GDR}$) and width ($\Gamma_{GDR}$) are the parameters that describe a GDR strength function. The knowledge in $E_{GDR}$ provides better understanding of the symmetry energy of nuclear matter while the systematic study of $\Gamma_{GDR}$ with $T$ sheds light on the characteristics of damping prevailing within the nuclear matter as well as the evolution of deformations embodied within it \cite{tri1}.   

Measurement of high energy $\gamma$-rays ($E_\gamma$ = 8 - 20 MeV) from the decay of hot compound nucleus (CN) is one of the most important probes for studying the GDR in excited nuclei. To understand the properties of the GDR parameters in excited nuclei, the characterization of the measured high energy $\gamma$-rays and its comparison with the predictions of theoretical statistical model related to CN decay are absolutely necessary. However, the acceptability of any statistical model prediction depends on its essential ingredient, the nuclear level density, which is also the central source of uncertainty in analyzing nuclear reactions and in reliable extraction of the GDR quantities i.e $S_{GDR}$, $E_{GDR}$ and $\Gamma_{GDR}$.

It is an important fact that the level density of excited nuclei is strongly influenced by the nuclear shell structure that melts down with the increase in excitation energy ($E^*$) of the nucleus. Although for high spin and high $E^*$ the shell structure might not be melted near the yrast line. The reliable extraction of the GDR Lorentzian not only depends on the statistical model predictions for the region $E_\gamma$ = 8-20 MeV of the high energy $\gamma$-ray spectrum but also on the statistical part ($E_\gamma \leq$ 8 MeV) of the spectrum. The statistical part is highly sensitive to the level density of the decaying nuclei, especially in the later part of the decay chain where shell effect is extremely important. Angular momentum ($J$) as well as the evolution of nuclear deformation in the CN decay chain also play leading roles in the modification of nuclear level density. Therefore, the statistical model must include a proper level density formalism as input, which can take care of all these facts.  
 
The basic nuclear level density formula, derived from the backshifted Fermi gas model and based on the pioneering work of Bethe \cite{bet1}, is given by 
\begin{equation}
\rho (E^*,J) = \frac{2J+1}{12 I ^{3/2}} \sqrt{a} \frac{\exp{(2\sqrt{(aU)}})}{U^2} \label{eqn1}
\end{equation}
where U = $E^*$-$\Delta$ - $J$($J$+1)$\hbar^2/2I$ is the available thermal energy. The effective moment of inertia of the compound nucleus is taken as $I = I_0 (1+\delta_1 J^2 + \delta_2 J^4)$, where $I_0$ is the spherical rigid body moment of inertia while $\delta_1$ and $\delta_2$ are the deformability coefficients. The excitation energy is back-shifted by the pairing energy $\Delta=12/\sqrt{A}$, $A$ being the mass number of the nucleus. $a$ is the level density parameter and is taken as an adjustable free parameter.

In the Fermi gas model the level density depends on the level density parameter $a$ which, in turn, is related to the finite size effect of the nuclear matter, the effective mass of the nucleon and the number of single particle levels near the Fermi surface. All of these depend on nuclear deformation, shell structure of the nucleus and also how the shell structure gradually melts with the increase in $E^*$ of the nucleus. P$\ddot{u}$hlhofer's \cite{pul1} statistical model code CASCADE includes formulation of level density parameter $a$ as per Dilg \cite{dil1} for $E^* < $ 10 MeV. For $E^* >$ 20 MeV, $a = A/k$ was used, based on the liquid drop model, where $k$ is user dependent free inverse level density parameter and $A$ is the nuclear mass. For $E^*$ ranging from 10 - 20 MeV, linear interpolation of $a$ and $\Delta$ is done in midway between the parametrization of Dilg and that of the liquid drop model. But the non-inclusion of the proper treatment of the shell corrections and its washing out at higher excitation energies along with the effect of nuclear deformation induces large uncertainty \cite{Kic87} in explaining the high energy $\gamma$-ray spectra when Dilg formulation is used. 

An ideal level density prescription should describe level density correctly starting from lower to higher $E^*$ at different $J$ values. It should also incorporate shell effects at lower $E^*$ smoothly connecting to the liquid drop behavior of the nucleus at higher $E^*$ and it must describe the high energy $\gamma$-ray spectra faithfully. The existing level density formulation by Ignatyuk, Smerekin and Tishin (IST) \cite{ign1} is quite popular as it can predict high energy $\gamma$-rays at different excitations. Besides IST, a few other modified theoretical as well as empirical level density prescriptions also exist in the literature. Two such level density prescriptions are of Kataria, Ramamurthy and Kapoor (KRK) \cite{kat1} and Budtz-Jorgensen and Knitter (BJK) \cite{bjk1}. 
Dioszegi et al \cite{dio1} have shown a comparative study between Dilg, IST and BJK level density formalisms for $A$ = 110 - 130 over the excitation energy range 58 - 62 MeV ($T$ = 1.98 - 2.23 MeV) and angular momentum 16.9-20.9 $\hbar$ by matching experimental high energy $\gamma$-ray spectra with CASCADE predictions. They pointed towards the superiority of IST level density over the others (Dilg and BJK) in the said $E^*$ region and also could not find any change of the GDR parameters even after including different level density prescriptions within the statistical model. However, they did not test the IST level density prescription at lower temperatures where the shell structure is important. Moreover, the other two prescriptions (KRK and BJK) have never been used in describing the high energy $\gamma$-ray spectra for different nuclei at different excitation energies.

In this work, the  KRK, BJK and IST level density formalisms are rigorously tested at lower excitation energy ranges $E^* \sim$ 29 - 50 MeV and lower spin (12 - 17 $\hbar$) for the CN $^{97}$Tc and $^{63}$Cu as well as at higher excitation energy $E^* \sim$ 109 - 121 MeV and higher spin (49 - 59 $\hbar$) for $^{113}$Sb. In addition, the applicability of IST and KRK level density prescriptions are also investigated on the high energy $\gamma$-ray spectra of $^{201}$Tl, in which the ground state shell correction energy is larger than $^{63}$Cu , $^{97}$Tc and $^{113}$Sb. Here, the advantage of populating these non-fissioning nuclei having ground state spherical is that the user-dependent free parameters in CASCADE decrease considerably and a one-component GDR strength function can be extracted  reliably. As a result, it becomes much easier to test the CASCADE predictions giving full emphasis only on the level density input. Moreover, we chose KRK level density rather than Dilg et al, as the former incorporates the shell structure of nuclei at lower $E^*$ and also its extrapolation to higher energies. 

\section {Experimental Details}
Very recently, a large amount of experimental data \cite{Sri08, supm11, dipu3, bala} on high energy $\gamma$-ray ($E_\gamma$ = 4 - 32 MeV) measurements have been reported in different nuclei and in varying excitation energies as well as angular momenta for studying the properties of the GDR modes in nuclei. Using the alpha beam from the K-130 cyclotron at the Variable Energy Cyclotron Centre, Kolkata, a self supporting 1 mg/cm$^2$ thick $^{93}$Nb target was bombarded at the projectile energies of 28, 35, 42 and 50 MeV populating the compound nucleus $^{97}$Tc at the excitation energies of 29.3, 36, 43 and 50.4 MeV, respectively \cite{bala}. The compound nuclei were populated in the angular momentum window 12 - 21 $\hbar$. The high energy $\gamma$-spectra were measured by means of a part (49 detectors in the form 7 $\times$ 7 matrix) of the LAMBDA array \cite{supm07}. The array was placed at a distance of 50 cm from the target and at an angle of 90$^0$ with the beam axis. The angular momentum populated by the compound nucleus was measured with a 50-element low energy $\gamma$-multiplicity detector array \cite{dipu5}. To measure the inverse level density  parameter (k) at different energies, the evaporated neutron energy spectra were extracted by converting the time of flight data of BC501A liquid scintillators \cite{kban1}. In other nuclear reactions $^{4}$He+$^{59}$Co, $^{20}$Ne + $^{93}$Nb, and $^{4}$He + $^{197}$Au, the compound nuclei $^{63}$Cu , $^{113}$Sb and $^{201}$Tl were populated for the beam energies of $E_{lab}$ = 35 MeV, 145 and 160 MeV, 42 and 50 MeV , respectively. The details of the experiments are explained elsewhere \cite{Sri08, dipu3}.  

\begin{figure}
\begin{center}
\includegraphics[height=5.0 cm, width=7.5 cm]{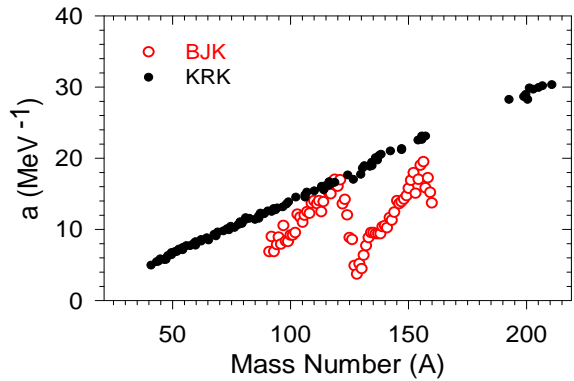}
\caption{\label{fig1} (Color online) The mass dependent level density parameters calculated by Kataria (i.e, KRK) for spherical nuclei (the filled circles). The red edged open circles are the level density parameters compiled by BJK from the experimental data on spontaneous fission of $^{252}Cf$.}
\end{center}
\end{figure}

\section{Adopted level densities}
The experimental spectra of high energy $\gamma$-rays coming from the decay of the compound nuclei ($^{63}$Cu, $^{97}$Tc, $^{113}$Sb and $^{201}$Tl) were fitted with the CASCADE predictions, using different level density prescriptions, folded with the detector response along with an exponential bremsstrahlung component given by $\exp(-E_\gamma/E_0)$ ($E_0$ is the slope parameter). The bremsstrahlung slope parameter $E_0$ was obtained from the sytematics $E_0 = 1.1[(E_{lab} - V_c)/A_p]^{0.72}$ where $E_{lab}$,$V_c$ and $A_p$ represent the beam energy, Coulomb barrier and projectile mass, respectively \cite{nif1}. Corresponding experimental angular momentum distribution, $E_{GDR}$, $S_{GDR}$ and $\Gamma_{GDR}$ were taken as CASCADE inputs. The GDR parameters were extracted by the $\chi^2$ best fit CASCADE predictions (in between $E_\gamma$ =10-20 MeV). For all the models, at a common beam energy, the bremsstrahlung slope parameter $E_0$ was kept fixed as per systematics and only the GDR strengths, widths and centroid energies were varied. 

\subsection{KRK level density prescription}
The semi-empirical model proposed by KRK \cite{kat1} on nuclear level density is important for statistical model of nuclear decay as it incorporates the shell effects and their $E^*$ dependence. The excitation energy and spin dependent level density $\rho$($E^*$, $J$), related with state density $W(E^*)$, adopted in this model is given by:

\begin{eqnarray}
\rho (E^*, J)=\frac{(2J+1)W(E^*)}{2\sqrt{(2\pi)} \sigma^3(E^*)} \exp{(\frac{-J(J+1)}{2\sigma^2(E^*)}})\\ 
W(E^*)=C\exp{S(E^*)}
\end{eqnarray}

where $\sigma$ is the spin cut-off parameter that depends on the effective moment of inertia of the nucleus. The state density $W(E^*)$ is related with entropy $S$, a function of excitation energy. $E^*$ and the temperature $T$ of the excited nucleus are interconnected by the level density parameter $a$. Unlike other level density formalisms, KRK propose the level density parameter as shell-independent similar to that of a nucleus under the liquid drop model. Here, the shell structure influences the level density with a ground state shell correction energy term added in the total nuclear $E^*$.

In KRK model, the analytical expressions of entropy and excitation energy can be achieved after detailed calculation as,

\begin {eqnarray} 
S=\frac{1}{3} \pi^2 g_0 T + \frac{A_1}{T} \frac{\pi^2 \omega ^2 \cosh {(\pi \omega T)}}{\sinh ^2 {(\pi \omega T)}}-\frac{\pi \omega T}{\sinh {\pi \omega T}}\\
E^*=\frac{1}{6} \pi^2 g_0 T^2 + A_1 (\frac{\pi^2 \omega ^2 T^2 \cosh{(\pi \omega T)}}{\sinh^2{(\pi \omega T)}})
\end {eqnarray}

where $A_1$, the ground state shell correction energy,  depends on the fundamental frequency of oscillation of the fluctuating part in level density ($\omega$).  At large temperature limit, one gets $S$=$\pi^2 g_0 T/3$ and $E^*$=-$A_1$+$\pi^2$g$_0$T$^{2}/6$, where $g_0$ is the density of single particle states proportional to the level density parameter $a$. These equations may be used as the framework to calculate the level density parameter as a function of $E^*$ with known values of ground state shell correction energies.
The constants $\alpha$, $\beta$ and $\omega_0$ can be estimated by comparing the theoretical nuclear level spacing with the experimental level spacing obtained from the neutron resonance data at $E^* \sim $ 10 MeV. These constants are related to the level density parameter and the frequency of shell oscillation ($\omega$) by $a$=$\alpha A (1-\beta A^{-1/3} B_s)$ and $\omega = \omega_0 A^{-1/3}$, where  $B_s$ is the surface area relative to that of a sphere of same volume. The reported best fit values were \cite{kat1}, $\alpha$ = 0.18 MeV$^{-1}$, $\beta$ = 1.0  and $\omega_0$ = 0.185 MeV$^{-1}$.  

KRK found that the level density calculated with this model successfully agrees with the experimental data up to the excitation energy around 25 MeV for the nuclei $^{56}Fe$ and $^{55}Mn$ \cite{hui1}. The mass dependent level density parameters calculated by KRK for spherical nuclei are shown in Fig \ref{fig1}. Unfortunately, the KRK model has not been used in the past for the extraction of GDR parameters.

\subsection{BJK level density prescription}

The second of the three nuclear level density prescriptions used in this work is that of BJK. The experimentally measured and compiled values of mass dependent level density parameters by BJK \cite{bjk1} for the nuclei ranging from 90 $\leq$ $A$ $\leq$ 165 are shown in Fig. \ref{fig1} by red edged open circles. These level density parameters were extracted from neutron evaporation measurements in spontaneous fission of $^{252}Cf$.

\subsection{IST level density prescription}

The model proposed by IST \cite{ign1} involves improved $E^*$-dependent level density parameter $a$. In this model the level density parameter is given by,  
\begin{equation}
a(U)=\tilde{a} (1+ \frac{\delta W}{U} (1-\exp{(-\gamma U)})
\end{equation}
This parametrization incorporates the effect of nuclear shell structure at lower excitation energy and extrapolates to the smooth liquid drop behavior at higher excitation energy where the shell effect is expected to be melted. $\delta W$ is the shell correction factor which is the difference between the experimental and the liquid drop masses. $\gamma^{-1}$ is the rate at which shell effects melt as $E^*$ increases and it is generally taken as 18.5 MeV. $\tilde{a}$ is asymptotic Fermi gas level density parameter and is taken as a user dependent free parameter. In partial modification of this formula, Reisdorf \cite{rei1} showed that asymptotic level density parameter depends on the mass of the compound nucleus as well as the nuclear deformation, given by: 
\begin{equation}
\tilde{a} = 0.04543r_0^3 + 0.1355 r_0^2 A^{-1/3} B_s + 0.1426 r_0 A^{-2/3} B_k
\end{equation}
where, $B_s$ and $B_k$  are the nuclear surface and curvature terms, respectively, and taken as 1 for spherical nuclei. $r_0$, the nuclear radius parameter, is taken as 1.15 fm.

\begin{figure}
\begin{center}
\includegraphics[height=8.0 cm, width=8.3 cm]{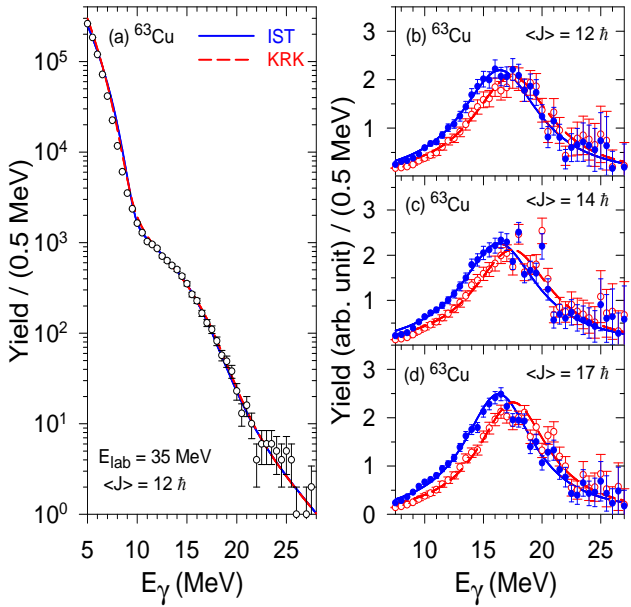}
\caption{\label{fig2} (Color online) Panel (a): The experimental high energy $\gamma$-ray spectra (open circles with error bars) for the reaction $^{4}$He+$^{59}$Co at projectile energy 35 MeV along with the CASCADE predictions utilizing KRK (red dashed line) and IST (blue continuous line) level density formalisms. Panels (b) to (d): The corresponding experimental linearized divided plots at different angular momenta IST (filled circles with blue continuous line) and KRK (open circles with red dashed line) level density prescriptions.}
\end{center}
\end{figure}

\begin{figure}
\begin{center}
\includegraphics[height=8.4 cm, width=7.8 cm]{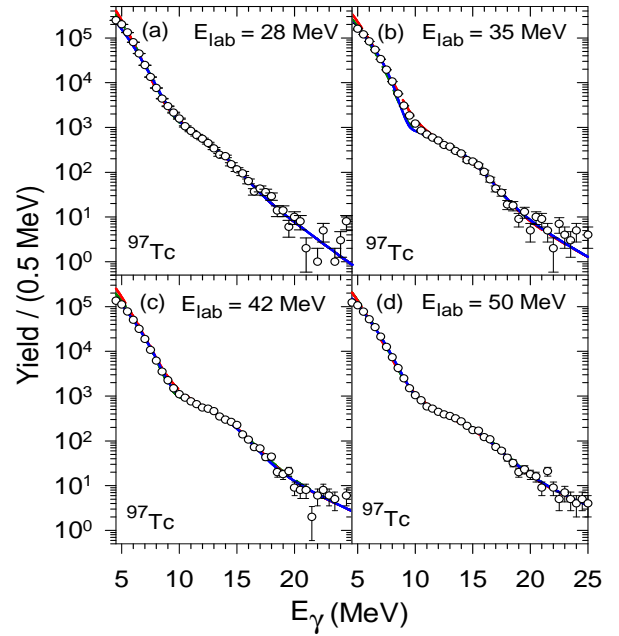}
\caption{\label{fig3} (Color online) Panels (a) to (d): The experimental high energy $\gamma$-ray spectra (open circles with error bars) for the reaction $^{4}$He+$^{93}$Nb at projectile energies 28, 35, 42 and 50 MeV along with CASCADE predictions utilizing KRK (red dashed line), BJK (green dotted-dashed line) and  IST (blue continuous line) level density formalisms.} 
\end{center}
\end{figure}

\begin{figure}
\begin{center}
\includegraphics[height=9.3 cm, width=8.3 cm]{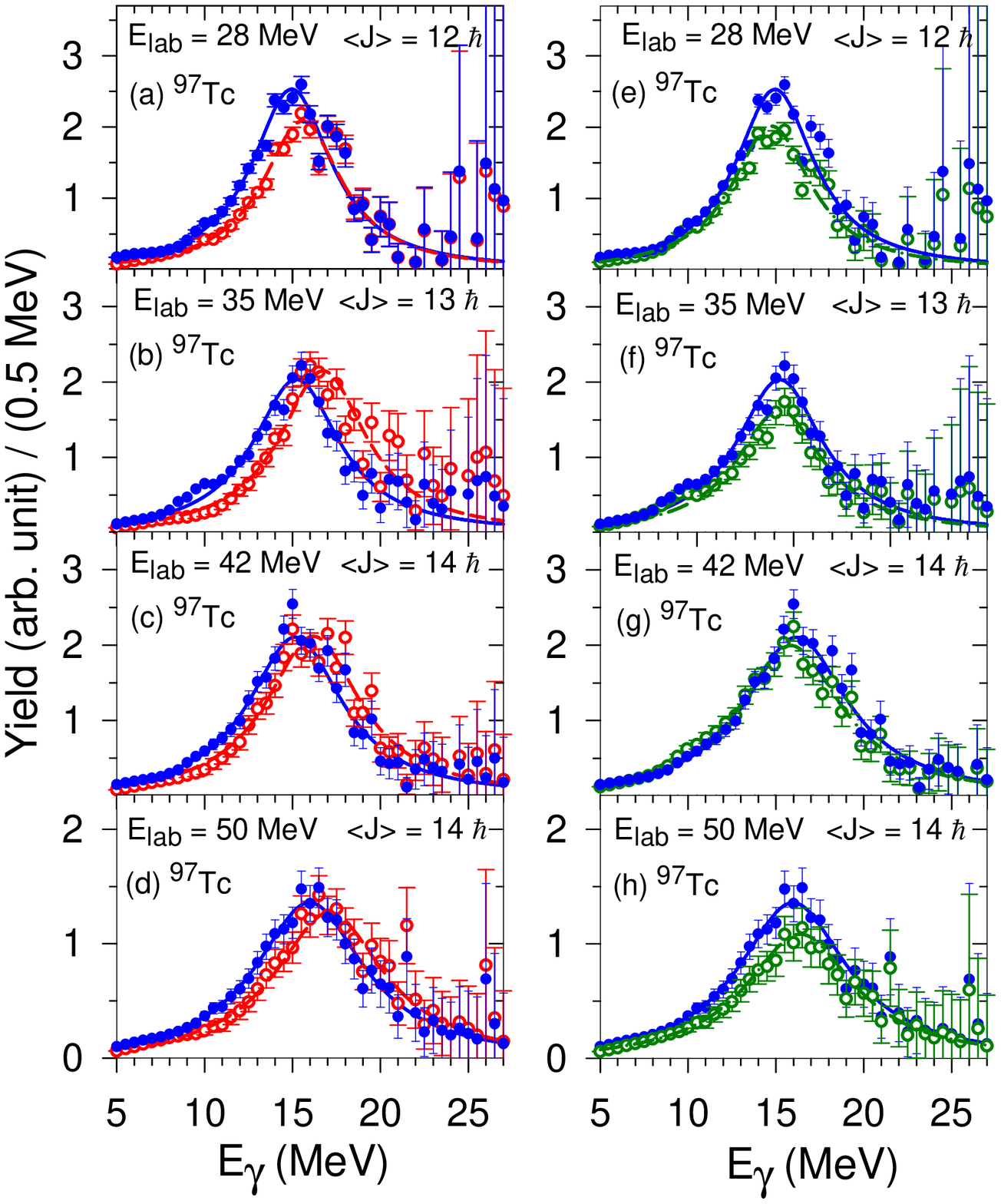}
\caption{\label{fig4} (Color online) Panels (a) to (d): The experimental linearized plots for the reaction $^{4}$He+$^{93}$Nb at projectile energies 28, 35, 42 and 50 MeV along with CASCADE predictions utilizing KRK (open circles with red dashed line) and  IST (filled circles blue continuous line) level density formalisms. 
Panels (e) to (h): Same as above but utilizing BJK (open circles with green dotted-dashed line) and  IST (filled circles with blue continuous line) level density formalisms.}  
\end{center}
\end{figure}

\begin{figure}
\begin{center}
\includegraphics[height=8.0 cm, width=8.4 cm]{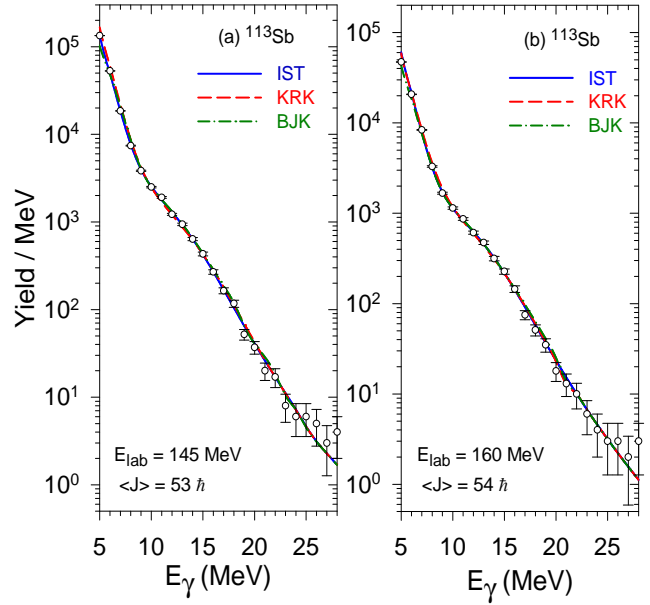}
\caption{\label{fig5} (Color online) (a) The experimental high energy $\gamma$-ray spectrum (open circles with error bars) for the reaction $^{20}$Ne + $^{93}$Nb at projectile energy 145 MeV along with CASCADE predictions exploiting KRK (red dashed line), BJK (green dotted-dashed line) and IST (blue continuous line) level density formalisms. (b) Same as (a) but for the reaction at projectile energy 160 MeV.}
\end{center}
\end{figure}

\begin{figure}
\begin{center}
\includegraphics[height=10.0 cm, width=8.4 cm]{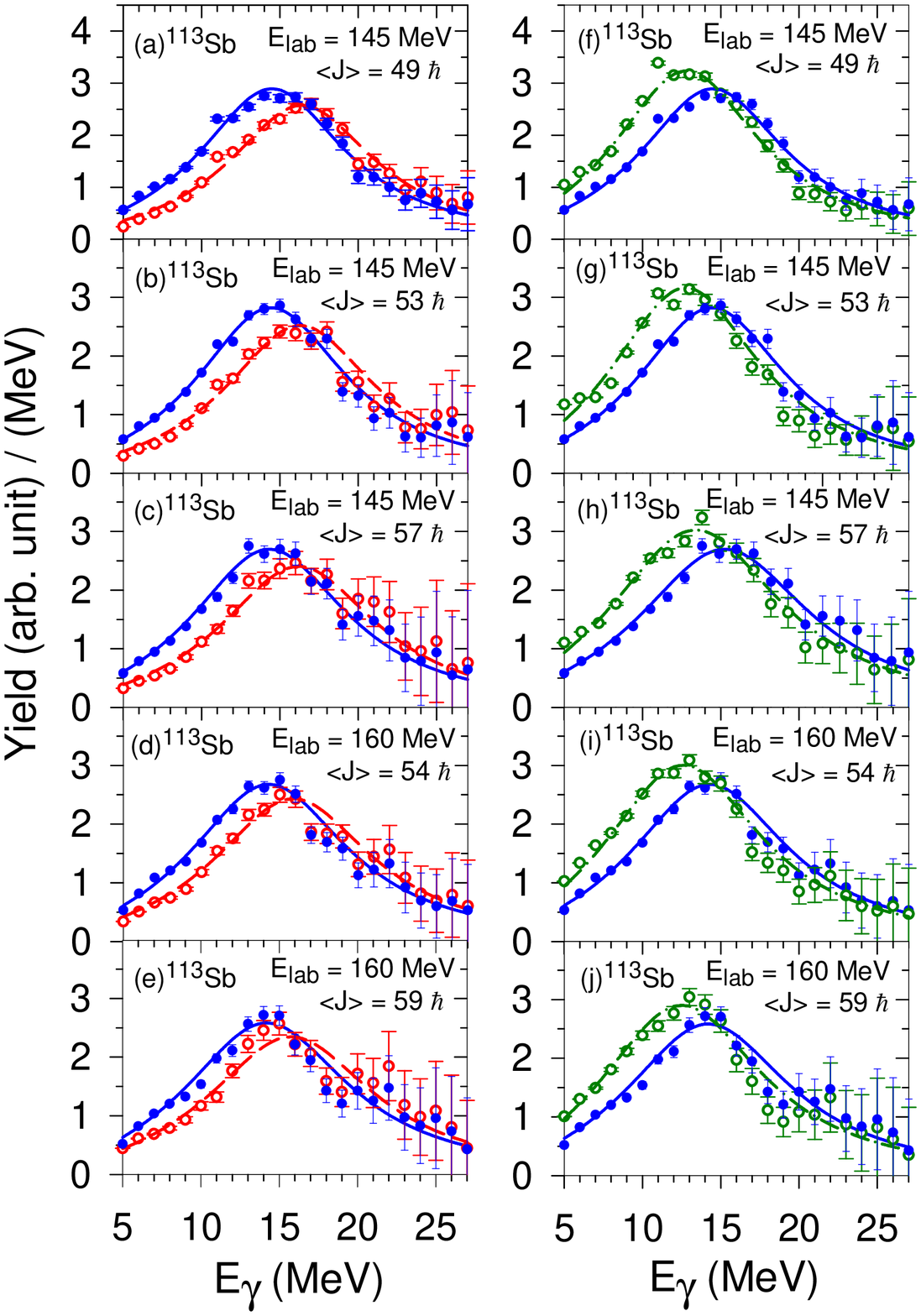}
\caption{\label{fig6} (Color online) Panels (a) to (e): The experimental linearized plots for the reaction $^{20}$Ne+$^{93}$Nb at projectile energies 145 and 160 MeV along with CASCADE predictions utilizing KRK (open circles with red dashed line) and  IST (filled circles with blue continuous line) level density formalisms. 
Panels (f) to (j): Same as above but utilizing BJK (open circles with green dotted-dashed line) and  IST (filled circles with blue continuous line) level density formalisms.}
\end{center}
\end{figure}

\begin{figure}
\begin{center}
\includegraphics[height=7.0 cm, width=8.4 cm]{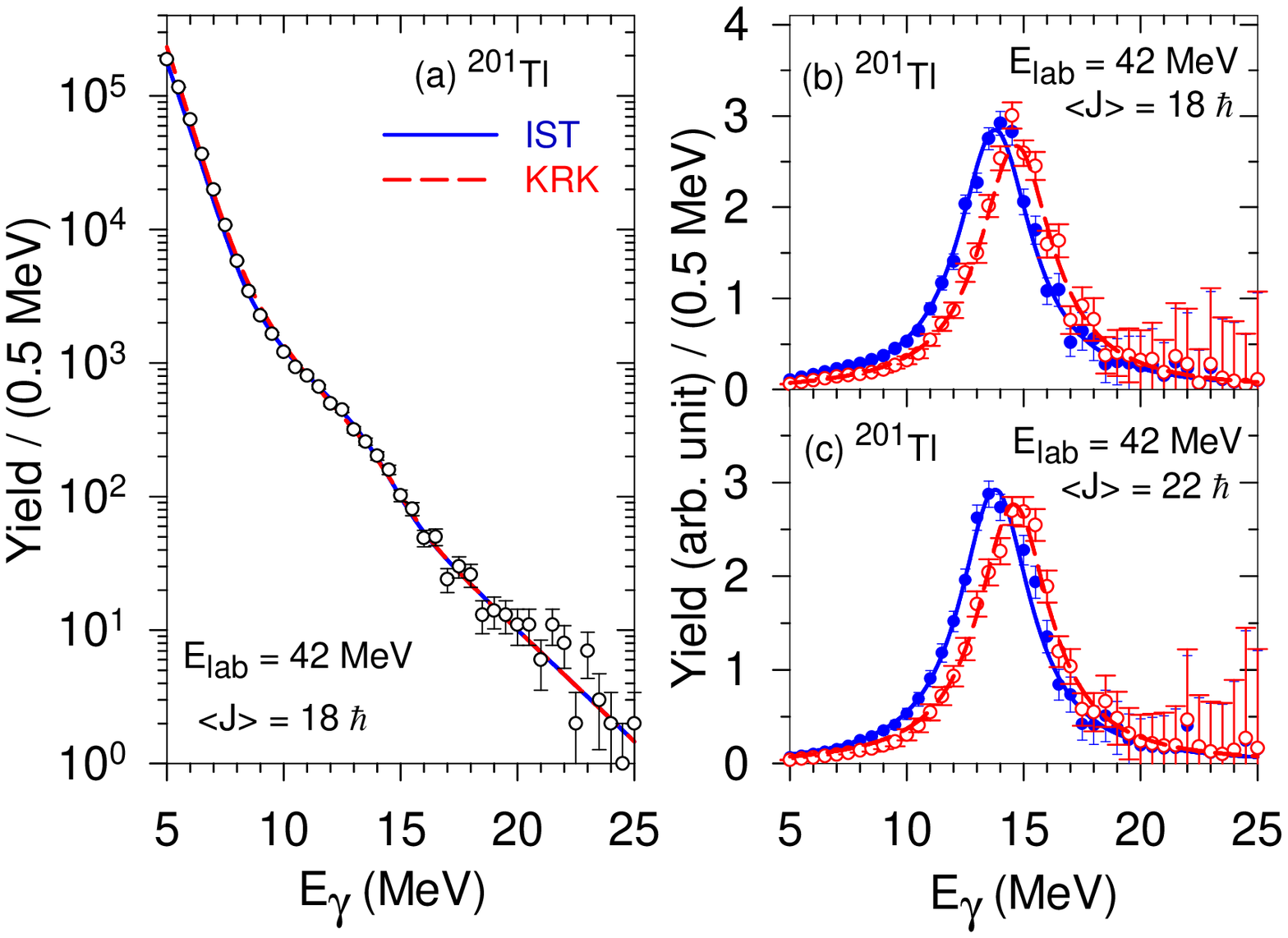}
\caption{\label{fig7} (Color online) Panel :(a) The experimental high energy $\gamma$-ray spectrum (open circles with error bars) for the reaction $^{4}$He + $^{197}$Au at projectile energy 42 MeV along with the CASCADE predictions exploiting KRK (red dashed line) and IST (blue continuous line) level density formalisms. Panels (b) and (c): The corresponding linearized plots.}
\end{center}
\end{figure}

\begin{figure}
\begin{center}
\includegraphics[height=7.0 cm, width=8.4 cm]{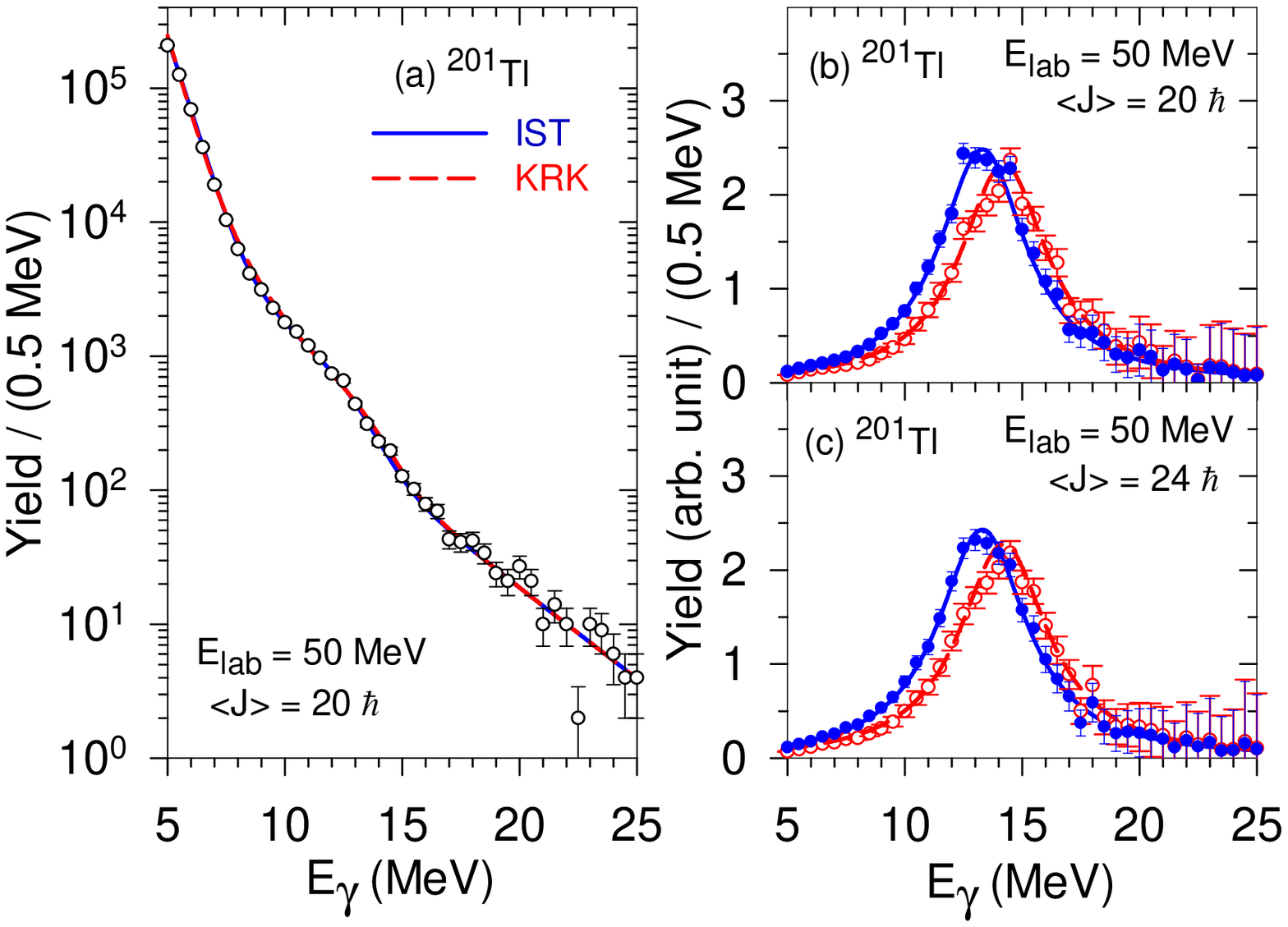}
\caption{\label{fig8} (Color online) Panel:(a) The experimental high energy $\gamma$-ray spectrum (open circles with error bars) for the reaction $^{4}$He + $^{197}$Au at projectile energy 50 MeV along with the CASCADE predictions exploiting KRK (red dashed line) and IST (blue continuous line) level density formalisms. Panels (b) and (c): The corresponding linearized plots.}
\end{center}
\end{figure}

\begin{figure}
\begin{center}
\includegraphics[height=8.0 cm, width=7.5 cm]{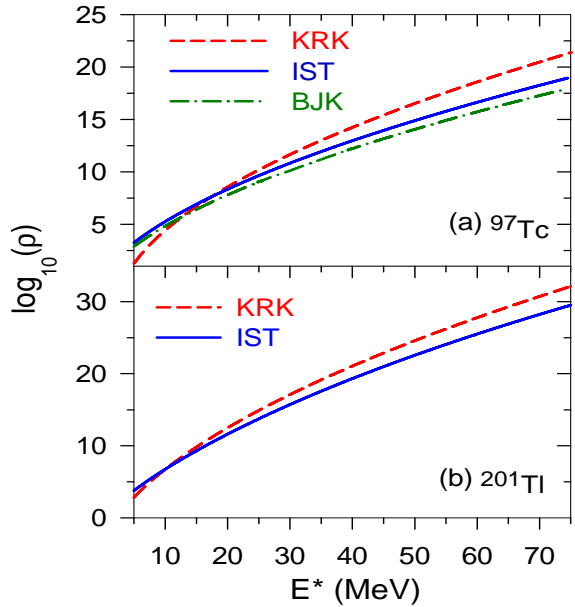}
\caption{\label{fig9} (Color online) Panel (a): The angular momentum integrated level density as a function of excitation energy $E^*$ for $^{97}$Tc. The blue continuous, green dotted-dashed and red dashed lines show the level densities due to IST, BJK and KRK prescriptions. Panel (b): Same as above but for 
$^{201}$Tl and considering IST and KRK prescriptions.}
\end{center}
\end{figure}

\begin{figure}
\begin{center}
\includegraphics[height=7.5 cm, width=8.4 cm]{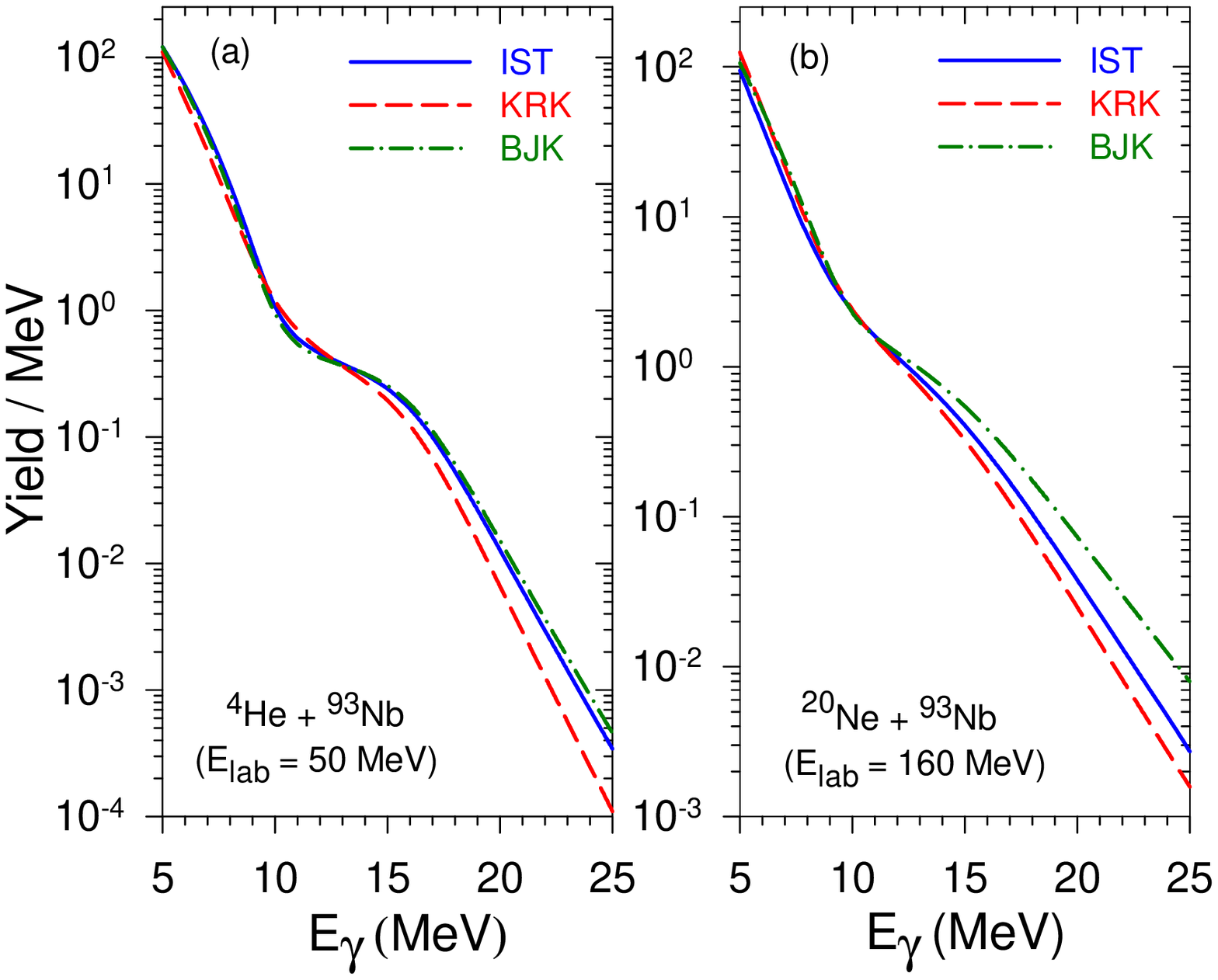}
\caption{\label{fig10} (Color online) Panel (a): CASCADE outputs of high energy $\gamma$-ray spectrum for the reaction $^{4}$He+$^{93}$Nb at projectile energy $E_{lab}$ = 50 MeV. The red dashed, blue continuous and green dotted-dashed lines correspond to the KRK, IST and BJK level density prescriptions, respectively. All the outputs are generated using same GDR parameters. Panel (b): Same as above but for $^{20}$Ne+$^{93}$Nb at projectile energy $E_{lab}$ = 160 MeV.}
\end{center}
\end{figure}

\section{Results and discussion}

In view of the available high energy $\gamma$-ray spectra measured by our group for the compound nuclei $^{63}$Cu, $^{97}$Tc, $^{113}$Sb and $^{201}$Tl over a wide range of  excitation energies 29.3 - 109 MeV and angular momenta 12-59 $\hbar$, we attempt here to assess the applicability of the three level density prescriptions. $^{63}$Cu, $^{97}$Tc and $^{113}$Sb, all of them have lower ground state shell correction energies (1.86 MeV, 1.26 MeV and 1.54 MeV, respectively). On the other hand, $^{201}$Tl has higher ground state shell correction energy of -8.27 MeV. While all the other three nuclei were populated at lower $E^*$ and $J$, $^{113}$Sb was populated at higher $E^*$ and $J$ values. 

\subsection{$^{63}$Cu}

The experimental high energy $\gamma$-ray spectrum for the compound nucleus $^{63}$Cu at projectile energy 35 MeV ($E^*$=36 MeV) 
and CN average angular momentum $J$=12 $\hbar$ is shown in panel (a) of Fig. \ref{fig2} by open circles. The CASCADE predictions utilizing IST (blue continuous line) and KRK (red dashed line) level density 
prescriptions are also included in the same figure.
It is highly interesting to note that both the IST and KRK level density formalisms included CASCADE represent 
the high-energy $\gamma$-ray spectra equally well. However, the extracted GDR centroid energies are very different. 
The discrepancy is evident in the linearized GDR plots shown in Fig.\ref{fig2}b-d for $J$=12, 14 and 17 $\hbar$ using 
the quantity $F(E_\gamma$)$Y^\textrm{exp}$($E_\gamma$)/$Y^\textrm{cal}$($E_\gamma$), 
where, $Y^\textrm{exp}$($E_\gamma$) and $Y^\textrm{cal}$($E_\gamma$) are the experimental 
and the best fit CASCADE spectra, corresponding to a single Lorentzian function F($E_\gamma$). 
The GDR centroid energy extracted at different $J$ using KRK prescription comes out to be 17.9 MeV. 
This value is slightly larger than the existing systamatics of GDR built on excited state: 
$E_{GDR}$ = 18 $A^{-1/3}$ + 25 $A^{-1/6}$ \cite{gaa1} which predicts 17.0 MeV. 

On the contrary, the IST level density included CASCADE successfully predicts the high energy $\gamma$-ray spectra 
for $^{63}$Cu with $E_{GDR}$ coming out between 16.7 to 16.9 MeV, much closer to the systematics. 
Unfortunately, it was not possible to test BJK level density for the nucleus $^{63}$Cu as lower 
mass $A$ = 63 does not fall under the mass distribution of $^{252}Cf$ fission fragment.

\subsection{$^{97}$Tc}

The experimental high energy $\gamma$-ray spectra for $^{97}$Tc at projectile energies of 28, 35, 42 and 50 MeV corresponding 
to $E^*$= 29.3, 36.0, 43.0 and 50.4 MeV are shown in the panels a-d of Fig. \ref{fig3} along with CASCADE 
predictions utilizing KRK (red dashed lines), BJK (green dotted-dashed line) and IST (blue continuous line) level density prescriptions. 
For better understanding of the GDR strength function, the corresponding linearized divided plots have also been shown in Fig. \ref{fig4}. 
The comparison between IST and KRK prescriptions is demonstrated in the panels (a)-(d) of the figure, 
while similar comparison between IST and BJK prescriptions is shown in the panels (e)-(h) of the same figure. 
The corresponding CN average $J$-values are also quoted in all the figures.

Here the linearized GDR Lorentzians once again corroborate the similar trend observed in $^{63}$Cu . The KRK prescription explains 
the GDR line shape well but with higher values of $E_{GDR}$.The extracted best fit $E_{GDR}$ is found to be lying between 17.0-17.5 MeV, 
much higher than the value of 15.6 MeV as per the existing systematics, except for $E_{lab}$ = 28 MeV in which the best fit value comes 
out to be 15.8 MeV closer to the systematics. It is worthwhile to mention that not only the CN $^{97}$Tc is populated in lower excitation 
energy ranges but also the compound nuclear angular momentum lies in lower side between 12 to 14 $\hbar$.    

The statistical model code CASCADE using BJK prescription can predict high energy $\gamma$-ray spectra reasonably well 
for $^{97}$Tc at all excitation energies. However, in contrast to KRK, the best fit GDR centroid energies 
vary between 15.0 MeV to 16.8 MeV.

All the experimental data for the decay of the CN $^{97}$Tc are found to be in good agreement with the 
CASCADE prediction utilizing IST level density prescription. The extracted best fit $E_{GDR}$ remains close 
to the value 15.6 MeV in agreement with the GDR systematics, except for $E_{lab}$ = 50 MeV in which the estimated 
GDR peak energy is around 16.4 MeV.  In all these data sets, the user dependent input $\tilde{a}$ is 
taken as $A/8.0$, $A/9.7$, $A/9.0$ and $A/9.2$ MeV$^{-1}$ at $E_{lab}$ = 28, 35, 42 and 50 MeV, respectively, as 
extracted from the neutron evaporation data \cite{bala}.

\subsection{$^{113}$Sb}

To understand the effect of level density prescriptions at higher angular momentum and higher excitation energy domains, the three level density prescriptions were used to explain the experimental data of high energy $\gamma$-rays measured for the compound nucleus $^{113}$Sb \cite{Sri08}. The experimental data (open circles) along with KRK (red dashed line), BJK (green dotted-dashed line) and IST (blue continuous line) predictions are shown in the panels (a) and (b) of Fig \ref{fig5} for $E_{lab}$=145 MeV ($E^*$=109 MeV) and average $J$=53 $\hbar$, $E_{lab}$=160 MeV ($E^*$=121 MeV) and average $J$=54 $\hbar$, respectively. The difference between KRK and IST predictions can be well understood through  linearized plots shown in the left panels (a)-(e) of Fig \ref{fig6} at $E_{lab}$=145 MeV and 160 MeV for different average CN angular momenta. Similarly, the right panels (f)-(j) of the same figure interpret the difference between IST and BJK predictions at similar projectile energies and average $J$. 
At higher $J$ (49-59 $\hbar$) values, for IST prescription, the asymptotic level density parameter was not measured and therefore $\tilde{a}$ is chosen as $A/8.0$ MeV$^{-1}$ according to Reisdorf formula \cite{rei1}. The change in $\tilde{a}$ from $A/8.0$ to $A/9.0$ MeV $^{-1}$ could only alter the extracted $E_{GDR}$ and $S_{GDR}$ by 3$\%$ and 5$\%$, respectively. For $^{113}$Sb, IST prescription explains the experimental data very well and $E_{GDR}$ = 15.5 MeV comes out to be consistent with the prediction(15.1 MeV) of the existing systematics. However, KRK prescription can explain the data only if $E_{GDR}$ is taken as 17.0-17.3 MeV, much larger than the existing systematics. BJK prescription can also explain the data but with much lower values of $E_{GDR}$ (14.0 MeV).

\subsection{$^{201}$Tl}

The suitability of three level density prescriptions were also investigated in higher nuclear mass region using another set of experimental data \cite{dipu3} for the reaction $^4$He  + $^{197}$Au at $E_{lab}$ = 50 MeV ($E^*$ = 47.5 MeV) and $E_{lab}$ = 42 MeV ($E^*$ = 39.5 MeV) at lower compound nuclear $J$ values (18-24 $\hbar$). The experimental high energy $\gamma$-spectra (open circles) are shown along with IST (blue continuous line) and KRK (red dashed line) prescriptions in the panels (a) of Fig. \ref{fig7} and Fig. \ref{fig8}. The corresponding linearized plots are shown in (b) and (c) of Figs. \ref{fig7} and  \ref{fig8}. Again similar trend has been found in which KRK predicted best fit $E_{GDR}$ comes out to be larger (14.4-14.7 MeV) than the excited state GDR centroid energy systematics (13.4 MeV), while IST predictions  of $E_{GDR}$ (13.5-13.9 MeV) remain in agreement with the systematics. In IST prescription, $\tilde{a}$ was taken as $A/8.0$ MeV$^{-1}$.
It was not possible to test BJK level density on the nucleus $^{201}$Tl as it does not fall under the mass distribution of $^{252}Cf$ fission fragment.

\subsection{Discussions}

It is highly interesting to note that the high energy $\gamma$-ray spectra for all the four nuclei  
$^{63}$Cu , $^{97}$Tc, $^{113}$Sb and $^{201}$Tl at different excitation energies and angular momenta 
are described very well using the KRK and IST level density formalism in the CASCADE calculation. 
BJK prescription can explain only $^{97}$Tc and $^{113}$Sb data as the other two nuclei do not fall under the 
mass distribution of $^{252}Cf$ fission fragments. However, it is important to mention that the extracted GDR centroid 
energies come out to be very dissimilar using different level density prescriptions but the GDR widths remain unchanged 
in all the cases. The extracted values of GDR parameters for the three prescriptions are shown in table \ref{tab1}. 

Despite testing on varying experimental data with wide range of $E^*$ $\&$ $J$ and ground state shell correction energies, 
intriguingly, the KRK prescription consistently predicts higher values of $E_{GDR}$. The $E_{GDR}$ for the 
nuclei  $^{63}$Cu , $^{97}$Tc and $^{113}$Sb with smaller ground state shell correction comes out to be 5, 13 and 12 $\%$ larger, respectively, than the existing systematics, 
while for $^{201}$Tl with larger ground state shell correction energy this discrepancy is nearly 10 $\%$. 
It is important to mention here that the KRK formalism is unique in comparison to the other two formalisms, 
as in this case the shell correction is incorporated through the nuclear excitation energy, instead of modifying 
the level density parameter. Figs. \ref{fig9}a and \ref{fig9}b show calculated $J$-integrated KRK level 
density (red dashed line) as a function of excitation energy $E^*$ along with the IST level density (blue continuous line) 
and BJK (green dot-dashed line) for the nucleus $^{97}$Tc (small shell correction) and $^{201}$Tl (large shell correction), 
respectively. As can be seen, KRK level density intersects IST at around $E^*$ = 20 MeV and thereafter 
they diverge from each other. This seems to be the possible reason of agreement between KRK and IST predicted 
data only at the effective $E^*$ (i.e $E^*$ of 29.3 MeV minus the rotational energy) $\sim$ 20 MeV for $^{97}$Tc. 
In case of higher $E^*$, the two prescriptions differ. As it appears, the larger GDR centroid energy obtained, 
in comparison to the systematic, for KRK prescription could be due to incorrect extrapolation at higher $E^*$ and $J$. 
In Fig. \ref{fig10} the CASCADE predictions of high energy $\gamma$-ray spectra have been shown utilizing 
KRK (red dashed line), IST (blue continuous line) and BJK (green dot-dashed line) level density prescriptions 
for the nuclei $^{97}$Tc at lower $E^*$ $\&$ $J$  and for  $^{113}$Sb at higher  $E^*$ $\&$ $J$. The GDR parameters 
were kept same for all the three prescriptions. The plots clearly indicate that for the common input parameters 
the KRK predicted $\gamma$-ray yield at higher energy side is lower compared to other two formalisms. 
This reduced yield is actually compensated by shifting $E_{GDR}$ at higher energies resulting higher 
values of $E_{GDR}$.  It is important to extract the correct centroid energy else it can introduce systematic 
error in the estimation of nuclear temperature for the GDR vibration. Apart from that a higher value of $E_{GDR}$ 
would jeopardize  the direct comparison between theoretical GDR lineshape and experimental GDR lineshape. 
Moreover, proper value of $E_{GDR}$ is also very useful in the exploration of nuclear symmetry energy \cite{tri1}. 
Hence, as it appears, the KRK level density prescription may not be used to explain the high energy $\gamma$-ray 
spectra as it systematically produces higher GDR centroid energy.    

It can be fairly inferred that the BJK prediction matches well with the experimental data for different $E^*$ 
values with reasonable best fit values of the GDR parameters, in so far as the nucleus $^{97}$Tc is concerned, 
which has a very small amount of ground state deformation as well as lower shell correction energy. However, 
at higher $J$ and higher $E^*$, BJK prescription misinterprets GDR centroid energies for $^{113}$Sb (Fig. \ref{fig9}).  
The inherent problem of BJK formalism lies in the fact that the compiled level density parameters are independent 
of $E^*$. Therefore, for same mass number, one has to adopt the same set of level density parameter for all values 
of $E^*$. This does not have much effect on $^{97}$Tc but has an adverse impact on the fitting of experimental 
spectra of $^{113}$Sb. BJK prediction breaks down not only at higher $E^*$, but also in the 
event of high angular momentum. The discrepancy at higher $E^*$ and $J$ is also highlighted in Fig. \ref{fig10}b. 
As can be seen, the BJK predicted $\gamma$-ray yield at higher energy side is much higher compared to other two 
formalisms for the common input GDR parameters. This higher yield is compensated by shifting $E_{GDR}$ at 
lower values. Another disadvantage concerned with BJK is that the extracted level density parameters from the neutron decay 
in $^{252}Cf$ fission studies can be used reliably only for a system with same mass and deformation which is 
available in the BJK compilations. Therefore we could not use the BJK prescription in case of $^{63}$Cu and $^{201}$Tl. 
Moreover, BJK level density cannot be safely extrapolated for other systems, especially in case of deformed 
nucleus as observed earlier \cite{dio1}. 

Interestingly, the IST level density formalism quite successfully describes the experimental data for all 
the four nuclei with reasonably correct values of $E_{GDR}$ at all conditions of $E^*$ $\&$ $J$ and ground state 
shell correction. Therefore, it can be commented that among the three level density prescriptions the IST 
is most suitable. The reason being its efficiency to interpret the varying experimental data sets in terms of 
best fit GDR parameters in agreement with existing systematics.

In recent times, several new semi-empirical as well as microscopic level density formalisms have also been developed. 
Nakada and Alhassid calculated level densities \cite{Naka} under the framework of Monte Carlo Shell Model for different nuclei. 
von Egidy and Bucurescu \cite{Egid} also estimated level densities using Fermi gas model as well as constant temperature model. 
It will be interesting in future, if the applicability of newer level density formalisms can also 
be tested on high energy $\gamma$-ray spectra.   

\section{Conclusion}
This work investigates the applicability of three different level density prescriptions viz. KRK, BJK and IST 
using experimental high energy $\gamma$-ray spectra for four nuclei $^{63}$Cu , $^{97}$Tc, $^{113}$Sb 
and $^{201}$Tl at different excitation energies and angular momenta. The extracted 
$E_{GDR}$ in case of KRK prescription was found to be higher than that of the existing GDR systematics 
for all the four nuclei owing to the prediction of larger values of the nuclear levels at higher excitation 
energies in comparison to those predicted by other two formalisms. On the other hand, the BJK prescription 
predicted lower $E_{GDR}$ compared to systematics in case of $^{113}$Sb at higher $E^*$ and $J$. Moreover, the 
BJK could not be tested on $^{63}$Cu and $^{201}$Tl due to its applicability in limited mass region. 
Intriguingly, the IST level density formalism quite successfully described all the data set both at low 
and high $E^*$ and $J$ indicating towards the universality of IST level density prescription in 
explaining the high energy $\gamma$-ray spectra with reasonably correct GDR parameters.

\begin{table}
\caption{\label{tab1} The GDR parameters extracted using different level density (l.d.) prescriptions.}  
\begin{tabular}{|c|c|c|c|c|c|c|c|c|c|c|c|c|c|}
\hline

CN     &  proj.  & $E_{lab}$ & $E^*$ &$J$$_{CN}$& \multicolumn{3}{|c|}{$E_{GDR}$ (MeV) for} & \multicolumn{3}{|c|}{$S_{GDR}$ for} & \multicolumn{3}{|c|}{$\Gamma_{GDR}$ (MeV) for}\\
        &             &  (MeV)        &(MeV)& $\hbar$  & \multicolumn{3}{|c|}{l.d. formalisms} & \multicolumn{3}{|c|}{l.d. formalisms} & \multicolumn{3}{|c|}{l.d. formalisms}\\  \hline
    
    &	       &		&	  &	    &	BJK  & KRK & IST & BJK	& KRK & IST & BJK & KRK & IST \\ \hline

	       &          &  35     &  36.0 & 12$\pm$6 &  & 17.9$\pm$0.1 &  16.9$\pm$0.1 & & 1.35 & 1.35& & 8.2 $\pm$0.2 & 8.2 $\pm$ 0.2\\
									
$^{63}$Cu     &  $^4$He   &  35     &  36.0 & 14$\pm$6 &   & 17.9$\pm$0.1 & 16.8 $\pm$0.1 & & 1.75 & 1.75& & 8.0 $\pm$0.2 
& 8.0 $\pm$ 0.2\\

	            &    &   35     &  36.0 & 17$\pm$6 &  & 17.9$\pm$0.1 &  16.7$\pm$0.1 & & 1.75 & 1.75& & 7.3 $\pm$0.2 & 7.3 $\pm$ 0.2\\ \hline

		          &         &   28    & 29.3 & 12$\pm$6 &  15.0$\pm$0.1  & 15.8$\pm$0.1 & 15.2$\pm$0.1 & 1.0& 1.0 & 1.20 & 5.5$\pm$0.5 &5.5$\pm$0.5 & 5.5$\pm$0.5\\             
 $^{97}$Tc   &  $^4$He &   35    & 36.0 &  13$\pm$4 &  15.5$\pm$0.1  & 17.0$\pm$0.1 & 15.6$\pm$0.1 & 1.1& 1.35 & 1.25 & 6.0$\pm$0.5 &6.0$\pm$0.5 & 6.0$\pm$0.5\\
              &         &   42    & 43.0 &  14$\pm$5 &  15.2$\pm$0.1  & 16.5$\pm$0.1 & 15.5$\pm$0.1 & 1.1& 1.30 & 1.20 & 6.5$\pm$0.5 &6.5$\pm$0.5 & 6.5$\pm$0.5\\
              &         &   50    & 50.4 &  14$\pm$5 &  16.8$\pm$0.1  & 17.5$\pm$0.1 & 16.4$\pm$0.1 & 0.9& 1.10 & 1.10 & 7.5$\pm$0.5 &7.5$\pm$0.5 & 7.5$\pm$0.5\\ \hline
							
              &          &   145   & 109.0 &  49$\pm$11 &  14.0$\pm$0.2  & 17.4$\pm$0.2 & 15.5$\pm$0.2 & 1.0& 1.0 & 1.0 & 11.6$\pm$0.3 &11.6$\pm$0.3 & 11.6$\pm$0.3\\ 
$^{113}$Sb    & $^{20}$Ne &  145   & 109.0 &  53$\pm$11 &  14.0$\pm$0.2  & 17.3$\pm$0.2 & 15.5$\pm$0.2 & 1.0& 1.0 & 1.0 & 11.8$\pm$0.3 &11.8$\pm$0.3 & 11.8$\pm$0.3\\    
              &          &   145   & 109.0 &  57$\pm$11 &  14.0$\pm$0.2  & 17.3$\pm$0.2 & 15.5$\pm$0.2 & 1.0& 1.0 & 1.0 & 12.4$\pm$0.3 &12.4$\pm$0.3 & 12.4$\pm$0.3\\ 

              &          &   160   & 121.0 &  54$\pm$11 &  14.0$\pm$0.2  & 17.0$\pm$0.2 & 15.5$\pm$0.2 & 1.0& 1.0 & 1.0 & 12.5$\pm$0.3 &12.5$\pm$0.3 & 12.5$\pm$0.3\\ 

	      &          &   160   & 121.0 &  59$\pm$11 &  14.0$\pm$0.2  & 17.0$\pm$0.2 & 15.5$\pm$0.2 & 1.0& 1.0 & 1.0 & 13.0$\pm$0.3 &13.0$\pm$0.3 & 13.0$\pm$0.3\\ \hline
                    
              &       &      42   & 39.5 &  18  $\pm$6 &         & 14.8 $\pm$0.3& 13.9$\pm$0.3 &  &1.0 & 1.0 &  &3.8$\pm$0.5 & 3.8$\pm$0.5\\ 
 $^{201}$Tl   &  $^{4}$He&   42   & 39.5 &  22  $\pm$6 &         & 14.7 $\pm$0.3& 13.9$\pm$0.3 &  &1.0 & 1.0 &  &3.7$\pm$0.5 & 3.7$\pm$0.5\\
              &          &   50   & 47.5 &  20  $\pm$6 &         & 14.4 $\pm$0.3& 13.5$\pm$0.3 &  &1.0 & 1.0 &  &4.5$\pm$0.5 & 4.5$\pm$0.5\\  
              &       &      50   & 47.5 &  24  $\pm$6 &         & 14.4 $\pm$0.3& 13.5$\pm$0.3 &  &1.0 & 1.0 &  &4.6$\pm$0.5 & 4.6$\pm$0.5\\ \hline                      
\hline
\end{tabular}
\end{table}

\end{document}